\documentclass[prb,twocolumn,aps,showpacs,fixfloats]{revtex4}
\usepackage{graphicx}
\usepackage{bm}
\usepackage{amsmath,amssymb}
\usepackage{subfigure}
\usepackage{float}
\usepackage{latexsym}
\usepackage{epstopdf}
\usepackage{color}
\usepackage{enumerate}
\usepackage{pdfpages}

\newcommand{\s}{\scriptscriptstyle}

\begin{document}

\title {Resonant magneto-tunneling
between normal and ferromagnetic  electrodes in relation to the three-terminal spin transport }

\author{Z. Yue  and M. E. Raikh }

 \affiliation{Department of Physics and
Astronomy, University of Utah, Salt Lake City, UT 84112, USA}

\begin{abstract}
The recently suggested mechanism [Y. Song and H. Dery, Phys. Rev. Lett. {\bf 113}, 047205 (2014)]  of the three-terminal spin transport
is based on the resonant tunneling of electrons between ferromagnetic and normal electrodes via an impurity. The sensitivity of current to a weak external magnetic field stems from a spin blockade, which, in turn, is enabled by strong on-site repulsion. We demonstrate that this sensitivity
 exists even in the absence of repulsion when a single-particle description applies. Within this description, we  calculate exactly the resonant-tunneling current between the electrodes. The
mechanism  of magnetoresistance,  completely different from the spin blocking,  has its origin in the interference of virtual tunneling amplitudes. Spin imbalance  in ferromagnetic electrode is responsible
for this interference and the resulting coupling of the Zeeman levels. This coupling also affects the current in the correlated regime.
\end{abstract}

\maketitle

\section{Introduction}
In the past decade there was a remarkable progress in fabrication
of lateral structures which combine ferromagnetic and normal layers and
exhibit spin transport. First experimental evidence of spin injection
from a ferromagnet into a nonmagnetic material was obtained with
the help of four-terminal (4T) technique. This technique was developed in the
pioneering papers Refs. \onlinecite{Silsbee1985},~\onlinecite{vanWeesPioneering}.
It utilizes two ferromagnetic electrodes, injector
and detector, coupled to a normal channel.  With  detector circuit being open,
the charge current does not flow between the electrodes. Instead,
the current circulating in the injector circuit leads to the
voltage buildup between the detector and the normal channel.
This nonlocal voltage is suppressed by a weak magnetic field
normal to the direction of magnetizations of the electrodes.
Such a suppression, called the
Hanle effect, reflects the precession of the spin of
carriers in course of diffusion between
the electrodes. Thus, the  characteristic width of the Hanle curve
is the inverse spin relaxation time.

\begin{figure}
\includegraphics[width=80mm]{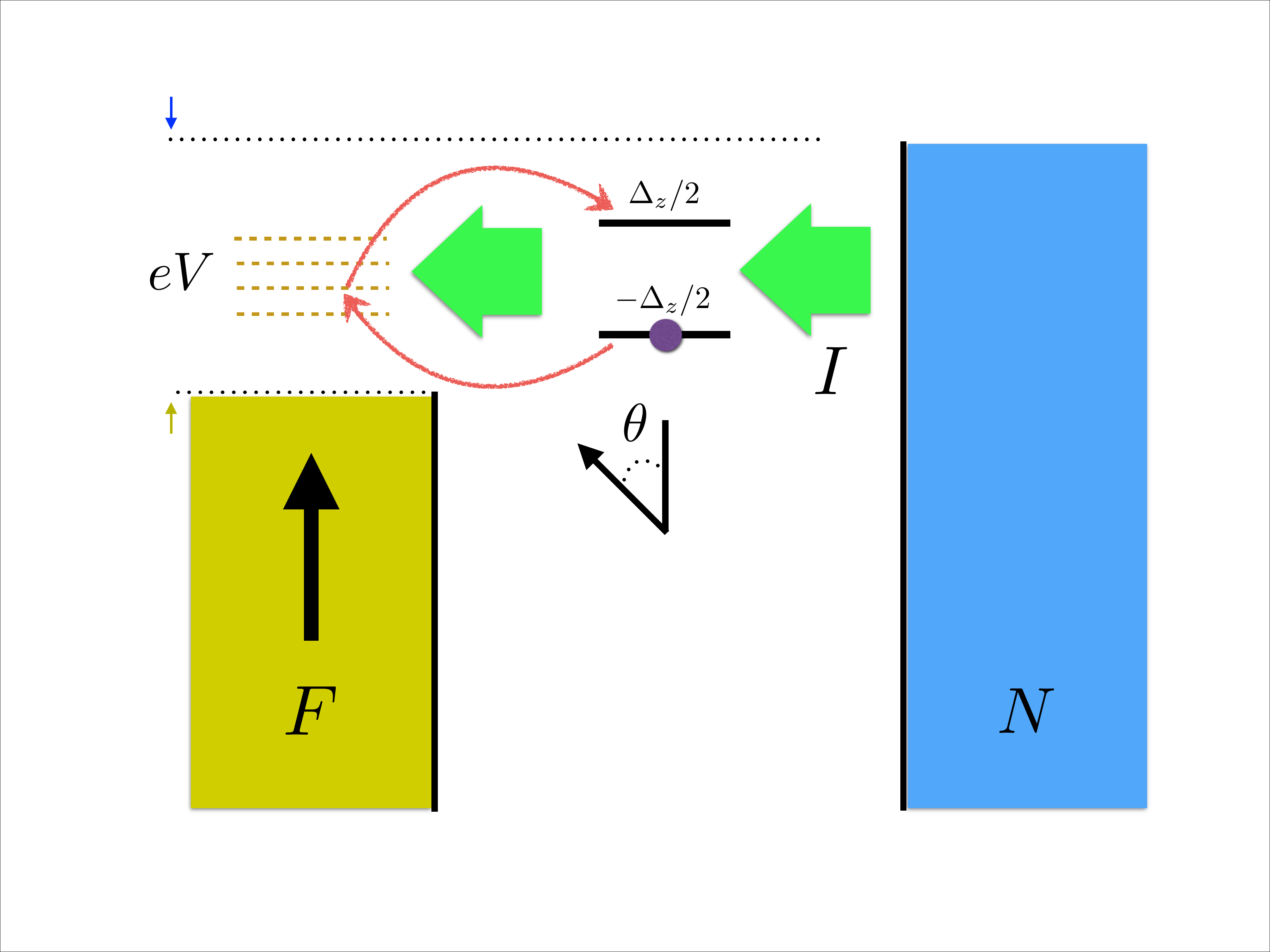}
\caption{ [Color online] Schematic illustration of resonant magnetotunneling between a normal electrode and a ferromagnet.
External field, tilted by an angle $\theta$ from the direction of magnetization, causes a splitting, $\Delta_z$, of the the impurity
level. For non-zero $\theta$ two Zeeman levels get coupled via a continuum of the states in a ferromagnet.}
\label{1}
\end{figure}

More recently,  experimental studies of  spin injection were carried out using the
three-terminal\cite{Jansen2009,Fert2009,Korean2011,Jansen2011,Tiwari,Jansen2012,Aoki2012,Jaffres2012,Jaffres'2012,Kasahara2012,Sinova2013}
(3T) technique.
Unlike the 4T technique,
in this technique the injector and detector electrodes are combined.
The signal measured is the contact voltage between the ferromagnet
and the normal channel. Sensitivity of this signal to the applied
magnetic field is simply the magnetoresistance.

Experimental results reported in Refs. \onlinecite{Jansen2009,Fert2009,Korean2011,Jansen2011,Tiwari,Jansen2012,Aoki2012,Jaffres2012,Jaffres'2012,Kasahara2012,Sinova2013} consistently  reveal two puzzling features of the 3T magnetoresistance.
Unlike the Hanle curves, the magnetoresistance
shows up for {\em both} orientations of the external
field parallel and perpendicular to the magnetization
of the injector. Moreover, the signs of magnetoresistance
are {\em opposite} for the two field orientations.
In addition, the 3T magnetoresistance curves are much broader
than the inverse spin-relaxation times measured independently.
In general, the basic underlying physics of magnetoresistance
in  transport between ferromagnetic and normal electrodes
constitutes a puzzle. Indeed, since the normal electrode, acting as a detector,
does not ``discriminate" between different spin orientations,
the current should not be sensitive to the spin precession.

Possible resolution of these puzzles was  proposed very recently
in the theoretical paper Ref. \onlinecite{Dery1}
and received some experimental support in the subsequent
publications Refs. \onlinecite{Stanford2014}-\onlinecite{Appelbaum2014}.
The main idea of Ref. \onlinecite{Dery1} is that the passage of current
between the ferromagnet and the
normal electrode can be modeled as resonant tunneling
via an impurity, see Fig. \ref{1}. On the qualitative level, the physics uncovered
in Ref. \onlinecite{Dery1} can be explained as follows.
When the current flows from normal into ferromagnetic electrodes,
the spins of electrons arriving on the impurity do not have a preferential
direction.  Suppose that the ferromagnet is fully polarized in
$\uparrow$ direction. Then electrons arriving with spin $\downarrow$
will never tunnel into the ferromagnet. External magnetic field induces
precession of spins of the arriving electrons. Then the electrons, which
were ``trapped" on the impurity without magnetic field, get a chance to
tunnel, unless the field is not parallel to the magnetization.
As a result, the current, which did not flow in a zero field, becomes
finite. Characteristic value of magnetic field can be estimated by
equating the period of precession to the waiting time for tunneling.
The mechanism is efficient if the spin relaxation rate is smaller than
the tunneling rate. Obviously, for the reverse bias, when electrons flow
from the ferromagnet this mechanism does not apply.

The key ingredient of the above scenario is a strong repulsion,
$U$, of $\uparrow$ and  $\downarrow$ electrons on the impurity. Indeed, if the
tunneling of the $\downarrow$ electron is forbidden, then, without the repulsion, the
current will be carried by $\uparrow$ electrons, so that there will be no ``blockade".

In the present paper we address a question: whether large $U$ is indeed necessary
to induce magnetoresistance. The question is delicate, since, for $U=0$, the current
does not depend on the polarity of bias. Thus, if magnetoresistance is finite for
tunneling into a ferromagnet, it should be the same for tunneling into a normal
electrode, which is highly non-obvious.
On the other hand, for $U=0$ the current can be calculated exactly.
Indeed, resonant tunneling in external field can be viewed as a
two-channel  resonant tunneling\cite{TwoChannel}
via the Zeeman-split levels. Our main analytical result is that
magnetoresistance is finite for $U=0$, and its magnitude is about
$50\%$. The physical origin of the magnetoresistance is the interference
of the two transport channels, or, in other words, the coupling
of Zeeman levels via a continuum of states in the ferromagnet.
We also trace how this coupling affects
the current in the regime of correlated transport\cite{Dery1}.

The paper is organized as follows. In Sect. II we derive and analyze the expression
for non-interacting resonant conductance via two Zeeman levels and, subsequently,
for the net resonant current. In Sect. III we study how the coupling of the Zeeman
levels via a ferromagnet affects the current in the presence of correlations.
Concluding remarks are presented in Sect. IV.

\begin{figure}
\includegraphics[width=80mm]{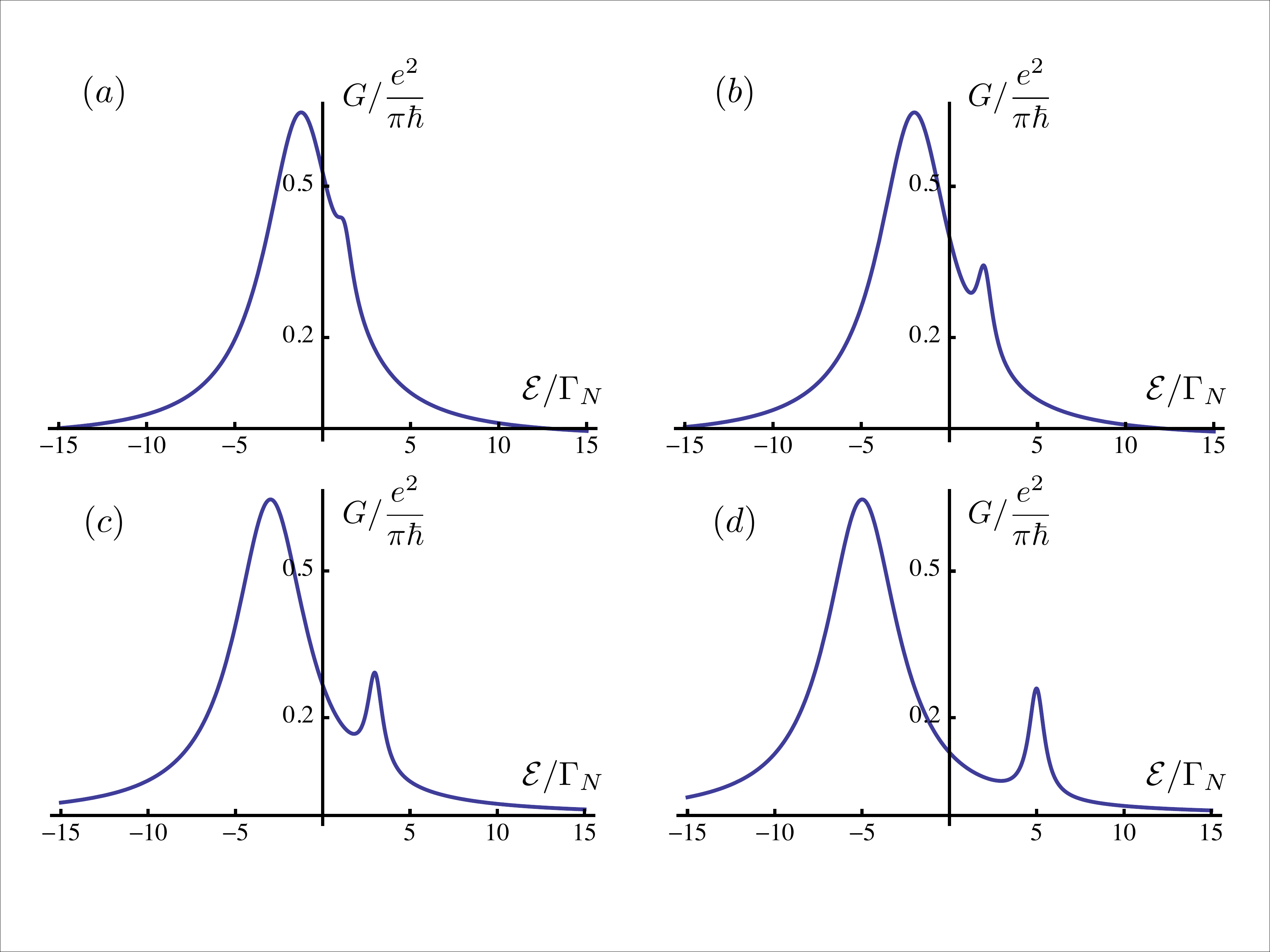}
\caption{ Differential conductance, $G({\cal E})$, in the units of $e^2/\pi\hbar$ is plotted from Eq. (\ref{Conductance})
for different dimensionless magnetic fields, in the units $\Delta_z/\Gamma_N$. (a)-(d) correspond to $\Delta_z/\Gamma_N = 2.5, 4, 6$, and $10$, respectively. All curves are plotted for $\Gamma_F=2\Gamma_N$ and the orientation of magnetic field, $\theta =15^\circ$.
 }
\label{2}
\end{figure}

\section{Magnetoresistance in the absence of Coulomb correlations}

\subsection{General expression}

Within a non-interacting picture  we can view the tunneling through a single impurity in a magnetic field as tunneling
via two Zeeman-split levels. The non-interacting current-voltage characteristics can be calculated from the tunnel conductance, $G({\cal E})$, as follows
\begin{equation}
\label{I}
I=\int d{\cal E} \left[f\Bigl({\cal E}-\frac{V}{2}\Bigr)-f\Bigl({\cal E}+\frac{V}{2}\Bigr)\right]G({\cal E}),
\end{equation}
where $V$ is the bias, and $f({\cal E})$ is the Fermi distribution.

If the electrodes are normal, the tunneling via each Zeeman level, $\pm \Delta_z/2$, proceeds independently, and  $G({\cal E})$ is given
by the Breit-Wigner formula
\begin{equation}
G_{\pm}({\cal E})=\frac{e^2}{\pi\hbar}\Bigl[\frac{\Gamma_L \Gamma_R}{({\cal E}\pm \frac{1}{2}\Delta_z)^2+ \frac{1}{4}(\Gamma_L+ \Gamma_R)^2}\Bigr],
\end{equation}
where $\Gamma_L$ and $\Gamma_R$ are the  widths with respect to tunneling into the left and right
electrodes, respectively.

Two tunneling channels are independent
because the normal electrodes do not couple the Zeeman levels,
since the corresponding spinors are orthogonal to each other.
By contrast,
a ferromagnetic electrode {\em does} introduce the
coupling between the levels for any  orientation of magnetic
field
except for the field parallel to the magnetization.
Indeed, if the angle between the magnetic field and
magnetization is $\theta$, the spinors corresponding
to the Zeeman levels are

\begin{equation}
\label{spinors}
\chi_+=\cos\frac{\theta}{2}\uparrow +\sin \frac{\theta}{2}\downarrow,~~~\chi_-=\sin\frac{\theta}{2}\uparrow -\cos\frac{\theta}{2}\downarrow,
\end{equation}
where $\uparrow$ and $\downarrow$ are the spin states in the ferromagnet,
and the azimuthal angle is set to zero. Denote with $\Gamma_L^{\uparrow}$ and $\Gamma_L^{\downarrow}$ the
widths of the Zeeman levels with respect to tunneling into the ferromagnet for $\theta=0$.
At finite $\theta$, an electron in the state $\chi_+$ can virtually tunnel into the \hspace{2mm} $\uparrow$-state of the
ferromagnet. The amplitude of this tunneling is $\cos\frac{\theta}{2}$. From the $\uparrow$-state it can then virtually
tunnel into $\chi_-$ with amplitude $\sin\frac{\theta}{2}$.  The electron can also proceed from $\chi_+$ to $\chi_-$
via the $\downarrow$ state of the ferromagnet. The corresponding amplitude is $-\sin\frac{\theta}{2}\cos\frac{\theta}{2}$,
 i.e. it has the opposite sign. As a result, the coupling matrix element between $\chi_+$ and $\chi_-$ is equal to
 $(\Gamma_L^{\uparrow}-\Gamma_L^{\downarrow})\sin\frac{\theta}{2}\cos\frac{\theta}{2}$. It is finite due to the difference
 in the densities of the intermediate states.

\begin{figure}
\includegraphics[width=80mm]{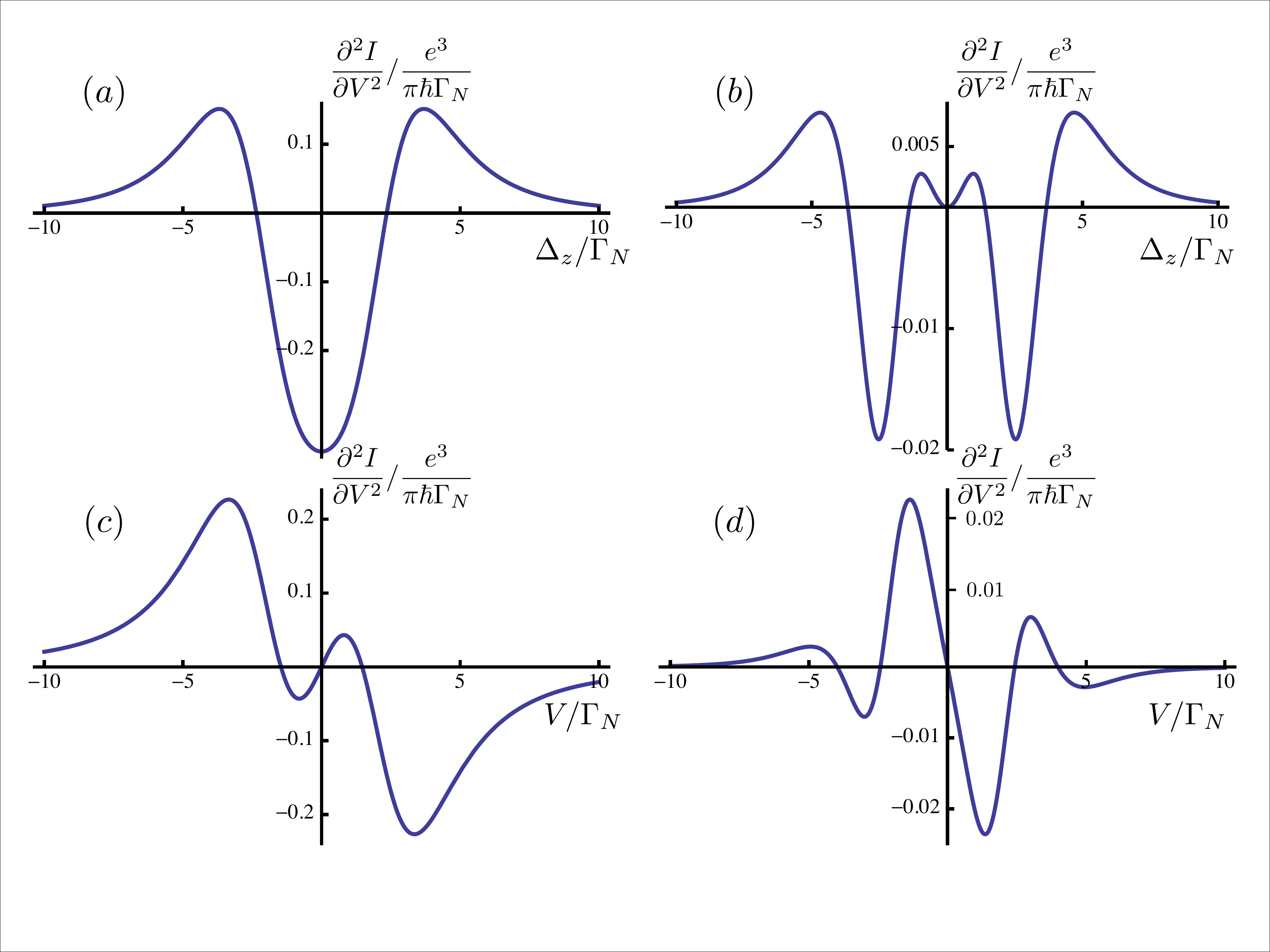}
\caption{ The second derivative, $\frac{\partial^2{I}}{\partial V^2}|_{\theta=\pi/2}$ ((a),(c)) and the difference,
$\frac{\partial^2{I}}{\partial V^2}|_{\theta=\pi/2}-\frac{\partial^2{I}}{\partial V^2}|_{\theta=0}$ ((b),(d))
is plotted in the  units $e^3/\pi\hbar\Gamma_N$ from Eqs. (\ref{I}), (\ref{Conductance}) versus dimensionless magnetic field, $\Delta_z/\Gamma_N$, (a) and (b), and versus dimensionless bias, $V/\Gamma_N$, (c) and (d). In (a) and (b) the bias is $V=2\Gamma_N$, while in (c) and (d) the magnetic field
is $\Delta_z=2\Gamma_N$. In all plots $\Gamma_F=1.5\Gamma_N$,  polarization is $p=1/3$, and temperature is $T=10\Gamma_N$.}
\label{3}
\end{figure}

\begin{figure}
\includegraphics[width=70mm]{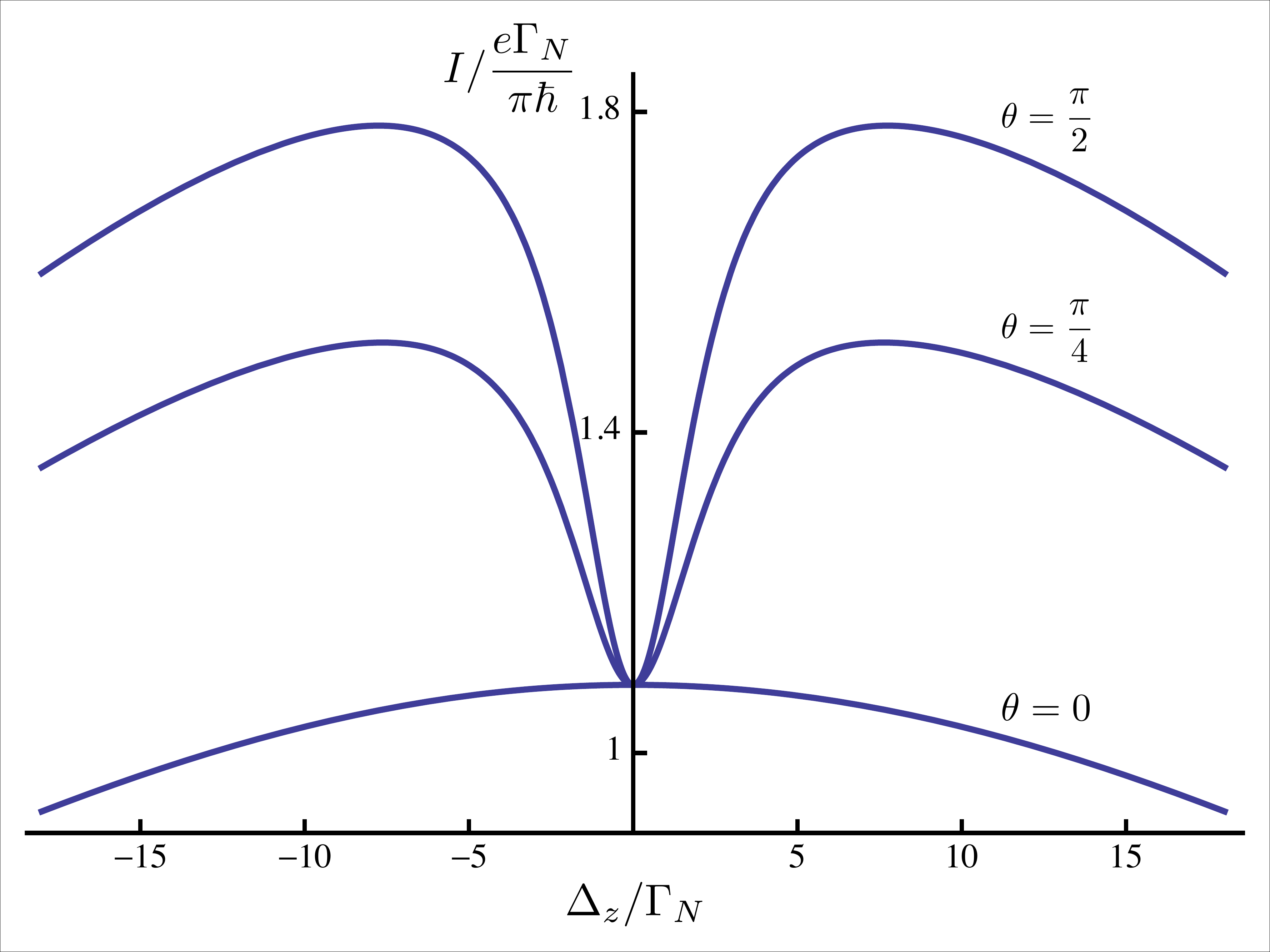}
\caption {Resonant current calculated numerically from Eqs. (\ref{I}), (\ref{Conductance}) is plotted  versus the dimensionless magnetic field, $\Delta_z/\Gamma_N$, for different field orientations. In all curves $\Gamma_F=2\Gamma_N$, the bias is $V=10\Gamma_N$ and the temperature is $T=10\Gamma_N$.
 }
\label{4}
\end{figure}

With two Zeeman levels coupled, the tunneling into the ferromagnet is described by a matrix
\begin{equation}
\label{matrix}
\hat{\Gamma}^L=\begin{pmatrix} \Gamma_L^\uparrow\cos^2\frac{\theta}{2}+\Gamma_L^\downarrow\sin^2\frac{\theta}{2} & (\Gamma_L^\uparrow-\Gamma_L^\downarrow)\cos\frac{\theta}{2}\sin\frac{\theta}{2} \\\\ (\Gamma_L^\uparrow-\Gamma_L^\downarrow)\cos\frac{\theta}{2}\sin\frac{\theta}{2}  & \Gamma_L^\uparrow\sin^2\frac{\theta}{2}+\Gamma_L^\downarrow\cos^2\frac{\theta}{2} \end{pmatrix}.
\end{equation}
This matrix enters into the calculation of the 
the differential conductance, which is given by the matrix generalization of the
Landauer formula
\begin{equation}
\label{G}
G({\cal E})=\frac{e^2}{\pi\hbar}{\text Tr}(\hat{\Gamma}^L \hat{S} \hat{\Gamma}^R \hat{S}^{\dagger} ),
\end{equation}
where 
the matrix  $\hat{\Gamma}^R$, describing the
the tunneling into the normal electrode, is diagonal
\begin{equation}
\label{Gamma}
\hat{\Gamma}^R=\Gamma_R \begin{pmatrix} 1 & 0\\0 & 1 \end{pmatrix}.
\end{equation}
The matrix $\hat{S}$, which is the  Green function in the matrix form, is given by
\begin{equation}
\label{S}
\hat{S}=\big[{\cal E}-\hat{E}+\frac{i}{2}(\hat{\Gamma}^L+\hat{\Gamma}^R)\big]^{-1}.
\end{equation}
The matrix $\hat{E}$ in Eq. (\ref{S}) encodes the energy level positions
\begin{equation}
\label{E}
\hat{E}=\begin{pmatrix} -\frac{1}{2}\Delta_z & 0\\0 & \frac{1}{2}\Delta_z \end{pmatrix}.
\end{equation}

We will present the result of the evaluating of the matrix product Eq. (\ref{G}) in the notations
of  Ref. \onlinecite{Dery1}, by denoting with $\Gamma_N$ (instead of $\Gamma_R$) the tunnel width for
the normal electrode and introducing the
of polarization, $p$, of the ferromagnetic electrode
\begin{equation}
p=\frac{\Gamma_L^{\uparrow}-\Gamma_L^{\downarrow}}{2\Gamma_F},
\end{equation}
where $\Gamma_F=\frac{1}{2}(\Gamma_L^{\uparrow}+\Gamma_L^{\downarrow})$ is the effective tunneling width for the
ferromagnetic electrode,
so that
\begin{equation}
\Gamma_L^{\uparrow}=\Gamma_F(1+p), ~~~~\Gamma_L^{\downarrow}=\Gamma_F(1-p).
\end{equation}
With the new notations, the matrix Eq. (\ref{matrix}) assumes a compact form
\begin{equation}
\label{matrix1}
\hat{\Gamma}^L=\Gamma_F \begin{pmatrix} 1+p\cos\theta & p\sin\theta\\p\sin\theta & 1-p\cos\theta \end{pmatrix}.
\end{equation}
The resulting expression  for conductance, $G({\cal E})$, reads

\begin{widetext}
\begin{equation}
\label{Conductance}
G({\cal E})=\frac{2e^2}{\pi\hbar}\Gamma_N\Gamma_F\frac{{\cal E}^2+\frac{1}{4}(\Delta_z^2+\Gamma_N^2)-{\cal E}\Delta_z p\cos\theta+\frac{1}{4}(1-p^2)\Gamma_F(2\Gamma_N+\Gamma_F)}
{\big[{\cal E}^2-\frac{1}{4}(\Delta_z^2+\Gamma_N^2+2\Gamma_F\Gamma_N+(1-p^2)\Gamma_F^2)
\big]^2+\big[{\cal E}(\Gamma_N+\Gamma_F)-\frac{1}{2}
\Gamma_F\Delta_z p\cos\theta\big]^2}.
\end{equation}
\end{widetext}
%

\subsection{Analysis}

Naturally, the dependence $G({\cal E})$ is an even function of ${\cal E}$
only for the perpendicular orientation of magnetic field, $\theta=\pi/2$.
The asymmetry of $G({\cal E})$ is maximal for the parallel orientation. The
asymmetry becomes progressively pronounced with increasing magnetic field,
as illustrated in Fig. \ref{2}.

The specifics of tunneling from the ferromagnet, as compared to the
normal electrode, is that Eq. (\ref{Conductance}) depends on the
orientation of magnetic field. In Ref. \onlinecite{Appelbaum2014}
the tunneling from cobalt-iron electrode into silicon via an oxide
was studied using the inelastic electron tunneling spectroscopy.
The curves $\frac{\partial^2{I}}{\partial V^2}$ exhibited different
behavior for parallel and perpendicular orientations of magnetic
field.    Motivated by this findings, in Fig. \ref{3} we plot the
 $\frac{\partial^2{I}}{\partial V^2}$ calculated from Eq. (\ref{Conductance})
for $\theta=\pi/2$ as a function of bias   and magnetic field together with
the difference of  $\frac{\partial^2{I}}{\partial V^2}$ for $\theta=\pi/2$
and $\theta =0$. The value at $\theta=0$ is finite due to finite polarization $p=1/3$ of the
ferromagnet.  All the  plots correspond to high temperature $T=10\Gamma_N$, so that only the magnitude, not the shape, of the curves is $T$-dependent. It is seen from Fig. \ref{3} that the relative difference of second derivatives  is $\sim 10\%$ and exhibits
additional structure at small $\Delta_z$ and at small bias. Still Fig. \ref{3} cannot account for the results of
Ref. \onlinecite{Appelbaum2014} where the observed anisotropy was really strong.

An interesting situation unfolds when the bias and temperature are of the same order and are much bigger than the level width.
Then the $\Delta_z$-dependence of current, calculated numerically from Eq. (\ref{I}), exhibits a growth for perpendicular orientation and a minimum
for parallel orientation as it is seen in Fig. \ref{4}.

\subsection{The net current at large bias}
In 3T experiments\cite{Jansen2009,Fert2009,Korean2011,Jansen2011,Tiwari,Jansen2012,Aoki2012,Jaffres2012,Jaffres'2012,Kasahara2012,Sinova2013}
 the net current showed the dependence on the magnitude and orientation of magnetic field.
It is not obvious whether this dependence is captured by Eqs. (\ref{I}), (\ref{Conductance}).
For tunneling between  normal electrodes, $p=0$, there should be no magnetoresistance. Indeed,
 the expression
Eq. (\ref{Conductance}) can be presented as a sum of two Lorentzians
\begin{multline}
\label{NetCurrent}
G({\cal E})=\frac{e^2}{\pi\hbar}\Gamma_N\Gamma_F\\
\times \Big[\frac{1}{({\cal E}- \frac{1}{2}\Delta_z)^2+ \frac{(\Gamma_F+ \Gamma_N)^2}{4}}
+\frac{1}{({\cal E}+ \frac{1}{2}\Delta_z)^2+ \frac{(\Gamma_F+ \Gamma_N)^2}{4}}\Big],
\end{multline}
so that the $\Delta_z$-dependence disappears upon integration over ${\cal E}$.
It turns out that magnetoresistance is nonzero for a finite $p$.
We will present the result for the net current assuming that ferromagnetic electrode
is fully polarized, $p=1$. Then the integration over  ${\cal E}$ yields
\begin{multline}
\label{current}
I(\Delta_z)=\frac{4e}{\hbar}\Gamma_F\Gamma_N \\
\times \frac{(\Delta_z^2+\Gamma_N^2+\Gamma_N\Gamma_F)
(\Gamma_N+\Gamma_F)-\Gamma_F\Delta_z^2\cos^2\theta}
{(\Delta_z^2+\Gamma_N^2+2\Gamma_N\Gamma_F)(\Gamma_N+\Gamma_F)^2
-\Gamma_F\Delta_z^2\cos^2\theta}.
\end{multline}
Eq. (\ref{current}) is our central result. Sensitivity of the net current  to $\Delta_z$
originates from the coupling of Zeeman levels via the ferromagnetic electrode [nondiagonal element in matrix Eq. (\ref{matrix1})] and, thus, it is most pronounced for $\Gamma_F \gg \Gamma_N$. In this limit Eq. (\ref{current}) can be simplified to
\begin{equation}
\label{GammaL>GammaR}
I(\Delta_z)=\frac{4e}{\hbar}\Gamma_N
\Biggl(1-\frac{\Gamma_N\Gamma_F}
{\Delta_z^2\sin^2\theta+2\Gamma_N\Gamma_F}\Biggr).
\end{equation}

\begin{figure}
\includegraphics[width=80mm]{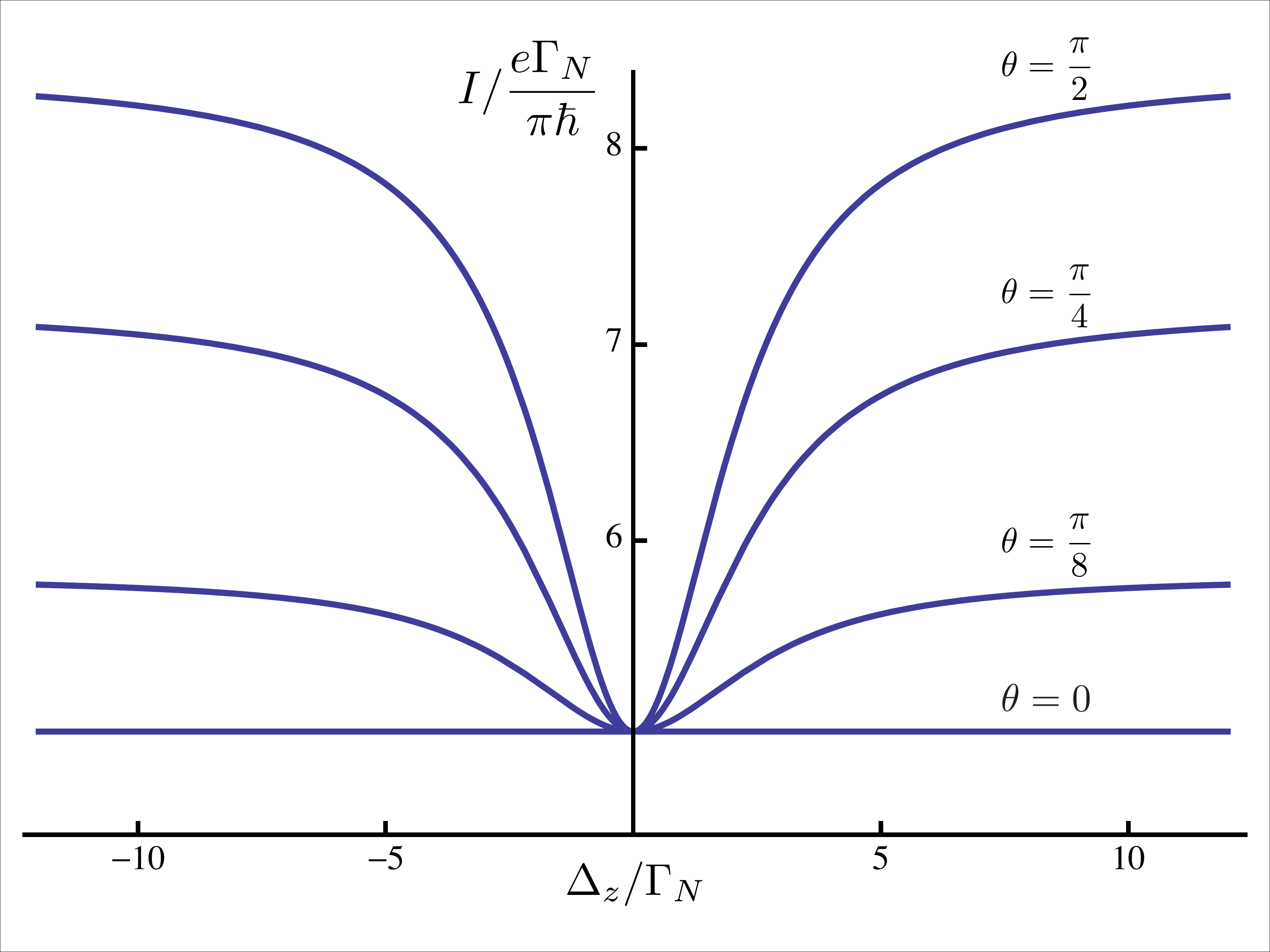}
\caption
 {Resonant current (in the units $e\Gamma_N/\pi\hbar$) in the absence of correlations is plotted from Eq. (\ref{current}) versus the dimensionless magnetic field, $\Delta_z/\Gamma_N$, for different field orientations. In all curves $\Gamma_F=2\Gamma_N$. }
\label{5}
\end{figure}

The current Eq. (\ref{current}) is a growing function of magnetic field for all orientations, $\theta$, see Fig. \ref{5}. At large $\Delta_z$
the current saturates at the value
\begin{equation}
\label{saturated}
I_{\infty}=\frac{4e}{\hbar}\Gamma_F\Gamma_N\frac{\Gamma_F\sin^2\theta+\Gamma_N}
{\Gamma_N^2+2\Gamma_F\Gamma_N+\Gamma_F^2\sin^2\theta}.
\end{equation}
This saturation value can be understood from the following argument.
At large $\Delta_z$ the tunneling via
upper and lower Zeeman levels get decoupled.
 The tunnel width of the upper level is $\Gamma_F\cos^2\frac{\theta}{2}+\frac{1}{2}\Gamma_N$, while the tunnel width of the lower level
is $\Gamma_F\sin^2\frac{\theta}{2}+\frac{1}{2}\Gamma_N$. The sum of the currents corresponding to these
widths yields Eq.~(\ref{saturated}). The same saturation value can be obtained from purely classical consideration, by introducing
the probabilities of all four variants of the occupation of the two Zeeman levels and solving the system of master equations for this probabilities.

It is quite nontrivial that in Eq. (\ref{GammaL>GammaR}) the characteristic scale of magnetic field, $\Delta_z\sim
\left(\Gamma_F\Gamma_N\right)^{1/2}$, is much smaller than the level width $\Gamma_F/2$. This suggests that, while
 the tunneling times for each of the Zeeman levels is $\Gamma_F^{-1}$, i.e. short,
 coupling of these levels via a ferromagnet modifies them in such a way that one of the resulting levels
 possesses a long lifetime.
 Similarly to Refs. \onlinecite{TwoChannel}, \onlinecite{SchultzVonOppen}, the origin  of this long lifetime can be traced to the complex poles of $G({\cal E})$ in Eq. (\ref{Conductance}). These poles correspond
to the condition: ${\text {det}}~ {\hat S}^{-1}=0$, where the matrix ${\hat S}$ is defined by Eq. (\ref{S}). The secular equation reads
\begin{multline}
\label{secular}
\Bigl[{\cal E}-\frac{1}{2}\Delta_z+\frac{i}{2}\left(\Gamma_N+(1-\cos\theta)\Gamma_F\right)\Bigr]\\
\times\Bigl[{\cal E}+\frac{1}{2}\Delta_z+\frac{i}{2}\left(\Gamma_N+(1+\cos\theta)\Gamma_F\right)\Bigr]=\Bigl(\frac{i}{2}\Gamma_F\sin\theta\Bigr)^2.
\end{multline}
The roots of Eq. (\ref{secular}) are
\begin{equation}
\label{Epm}
{\cal E}_{\pm}=\frac{1}{2}\Bigl[i(\Gamma_N+\Gamma_F)\pm \sqrt{\Delta_z^2-\Gamma_F^2-2i\Delta_z\Gamma_F\cos\theta}~\Bigr].
\end{equation}
For $\Gamma_N \ll \Gamma_F$ and $\Delta_z \ll \Gamma_F$ the imaginary parts of the roots are
\begin{equation}
{\text {Im}}~{\cal E}_1=\frac{1}{2}\Gamma_F,~~~~~
{\text {Im}}~{\cal E}_2=\frac{\Gamma_N}{2}+\frac{\Delta_z^2\sin^2\theta}{4\Gamma_F}.
\end{equation}
We see that the time $({\text {Im}}~{\cal E}_2)^{-1}$ is long, and defines the scale $\Delta_z\sim
\left(\Gamma_F\Gamma_N\right)^{1/2}$ of magnetic field.


\begin{figure}
\includegraphics[width=80mm]{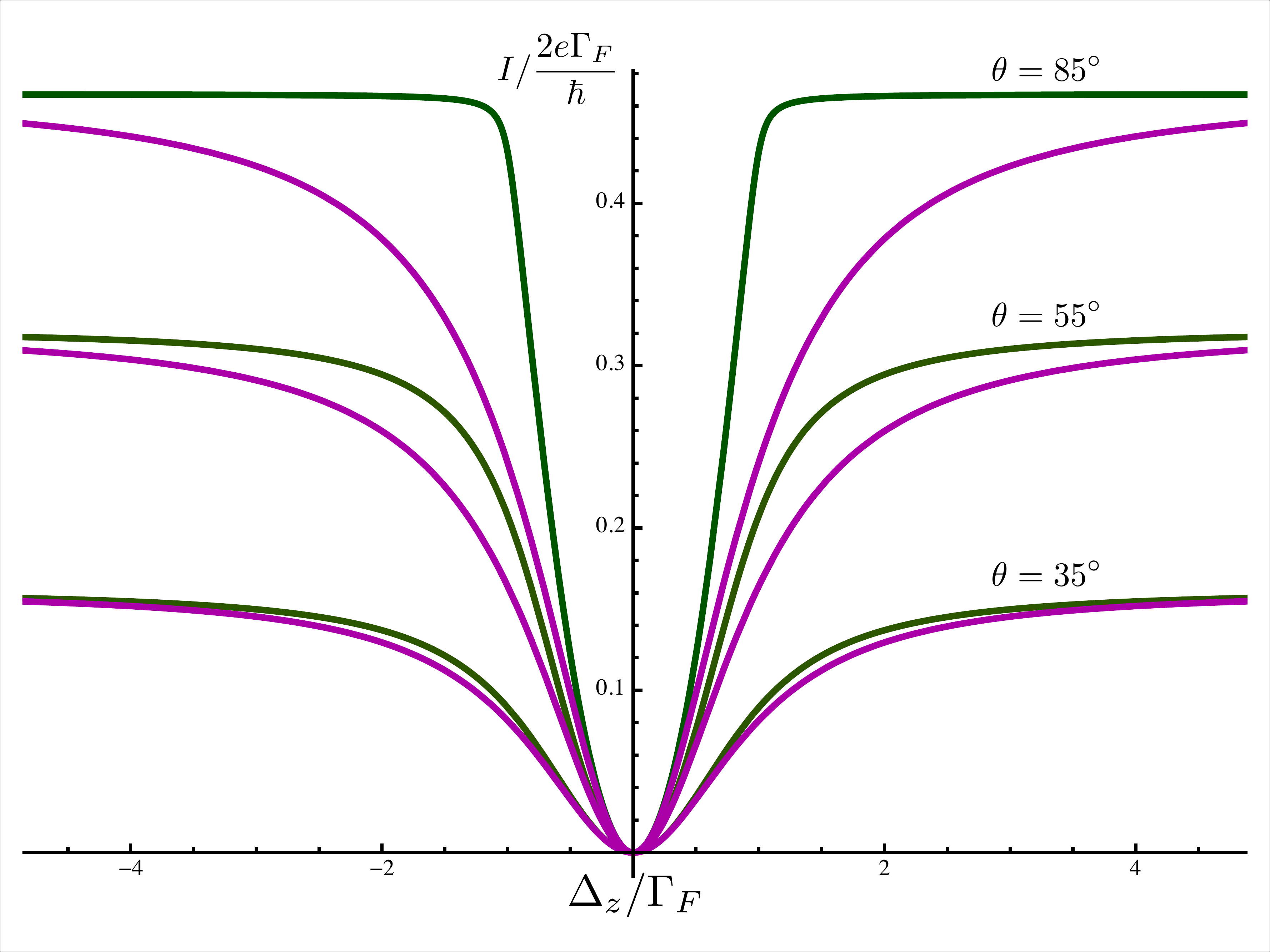}
\caption{[Color online] The current from normal into ferromagnetic electrode (in the units $2e\Gamma_F/\hbar$) in the correlated regime is plotted versus dimensionless
magnetic field, $\Delta_z/\Gamma_F$, for different orientations, $\theta$ and $\Gamma_N=8\Gamma_F$. Green curves are plotted from
Eq. (\ref{result}), while the purple curves are plotted from Eq. (\ref{Deryresult}), Ref. \onlinecite{Dery1}.
 }
\label{6}
\end{figure}

\section{Correlated tunneling}

 With strong on-site repulsion, $U$, and the bias, $V$, exceeding the Kondo temperature, the mechanism of transport is sequential tunneling. The scenario of this sequential tunneling is most simple for $U \gg V$, when the double occupancy
 of the impurity is forbidden. Then the passage of current, say, from the ferromagnet $(F)$ into normal electrode $(N)$
 via impurity $(i)$ proceeds in simple cycles. At the first step,
 the electron tunnels from $F $ to $i$, and at the second step, from $i$ to $N$.
 The current is inversely proportional to the average duration of the cycle, i.e.
 \begin{equation}
 \label{F-N}
 I^{F\rightarrow N}=\frac{e}{\tau^{F\rightarrow i}+\tau^{i\rightarrow N}},
 \end{equation}
 where $\tau^{F\rightarrow i}$  and $\tau^{i\rightarrow N}$ are the {\em average} waiting times for the
 corresponding tunneling processes.
 Similarly, the current from $N$ to $F$ is given by
\begin{equation}
 \label{N-F}
 I^{N\rightarrow F}=\frac{e}{\tau^{N\rightarrow i}+\tau^{i\rightarrow F}}.
 \end{equation}

For a normal electrode, the time  $\tau^{i\rightarrow N}$ is related to  $\tau^{N\rightarrow i}$ as\cite{GlazmanMatveev2}
\begin{equation}
\tau^{i\rightarrow N}=2\tau^{N\rightarrow i},
\end{equation}
reflecting the fact that tunneling from the electrode onto an empty impurity is possible for both spin directions, while
the electron on impurity can tunnel only into the states in the electrode having the same spin direction.
If the electrode $F$ was unpolarized, the currents Eqs. (\ref{F-N}) and (\ref{N-F}) would be given by\cite{GlazmanMatveev2}
 \begin{eqnarray}
 \label{TwoCurrents}
 I^{F\rightarrow N}=\frac{e}{2\tau^{i\rightarrow F}+\tau^{i\rightarrow N}},\\
 I^{N\rightarrow F}=\frac{e}{\tau^{i\rightarrow F}+2\tau^{i\rightarrow N}}.
 \end{eqnarray}

For a polarized electrode $F$ the relation  $\tau^{i\rightarrow F}=2\tau^{F\rightarrow i}$ is not valid.
In calculating $\tau^{i\rightarrow F}$ one should keep in mind that electron can tunnel into $F$ {\em from}
both Zeeman levels described by spinors $\chi_+$, $\chi_-$, Eq. (\ref{spinors}), so that
\begin{equation}
\label{iF}
\tau^{i\rightarrow F}=\frac{\tau_+^{i\rightarrow F} +\tau_-^{i\rightarrow F}}{2}.
\end{equation}
In the same way, in calculating $\tau^{F\rightarrow i}$, it should be taken into account that the electron from
$F$ can tunnel {\em into} both Zeeman levels. The net rate of tunneling is given by
\begin{equation}
\label{Fi}
(\tau^{F\rightarrow i})^{-1}=(\tau_+^{F\rightarrow i})^{-1}+(\tau_-^{F\rightarrow i})^{-1}.
\end{equation}
Upon these modifications, the times $\tau^{i\rightarrow F}$ and $\tau^{F\rightarrow i}$ can be very different.
Suppose that the polarization is full, $p=1$, and that the magnetic field is directed along the direction of magnetization.
Then for $\tau_+^{i\rightarrow F}$ we have $\tau_+^{i\rightarrow F}=(2\Gamma_F)^{-1}$, while $\tau_-^{i\rightarrow F}=\infty$, 
reflecting the fact\cite{Dery1} that electron with spin $\downarrow$ cannot tunnel into $F$, where all spins are $\uparrow$.
For a finite angle, $\theta$, between magnetization and magnetic field this blockade is lifted.

In calculating the tunneling times, it is very important
that electron tunnels into $F$ not from {\em pure} Zeeman levels, but from the levels coupled via $F$. This coupling is described
by the non-diagonal element of the matrix Eq. (\ref{matrix1}). Then the corresponding partial times are given by\cite{TwoChannel}
 \begin{eqnarray}
 \label{Tau+-}
 \tau_+^{i\rightarrow F}=\tau_+^{F\rightarrow i}=\frac{\hbar}{2\text{Im}~ {\cal E}_+},\\
 \tau_-^{i\rightarrow F}=\tau_-^{F\rightarrow i}=\frac{\hbar}{2\text{Im}~  {\cal E}_-},
 \end{eqnarray}
where ${\cal E}_+$ and ${\cal E}_-$ are given by Eq. (\ref{Epm}) with $\Gamma_N=0$.

It can be easily seen from Eq. (\ref{Epm}) that the relation
\begin{equation}
\tau_+^{F\rightarrow i}+\tau_-^{F\rightarrow i}=\frac{\hbar}{2\Gamma_F}
\end{equation}
holds. This, in turn, means that the current $I^{F\rightarrow N}$ is simply equal to $\frac{2e}{\hbar}\Gamma_F\Gamma_N/(2\Gamma_F+\Gamma_N)$,
i.e. it does not exhibit any magnetic-field dependence\cite{Dery1}.
On the other hand, with times given by Eq. (\ref{Tau+-}), the current $I^{N\rightarrow F}$ acquires a non-trivial $\Delta_z$-dependence. Taking into account that

\begin{multline}
\label{E+-}
{\text{Im} {\cal E}_\pm}=\\ \frac{\Gamma_F}{2}\pm\frac{1}{2}
\Bigl[-\frac{1}{2}(\Delta_z^2-\Gamma_F^2)+\frac{1}{2}
\sqrt{(\Delta_z^2-\Gamma_F^2)^2+4\Gamma_F^2\Delta_z^2\cos^2\theta}~\Bigr]^{\frac{1}{2}}      ,
\end{multline}
we get

\begin{multline}
\label{result}
I^{N\rightarrow F}= \Bigl(\frac{2e}{\hbar}\Bigr)\\
\times \frac{\Gamma_N\big(\Delta_z^2+\Gamma_F^2-\sqrt{(\Delta_z^2-\Gamma_F^2)^2+4\Gamma_F^2\Delta_z^2\cos^2\theta}~\big)}
{4\Gamma_N\Gamma_F+\Delta_z^2+\Gamma_F^2-\sqrt{(\Delta_z^2-\Gamma_F^2)^2+4\Gamma_F^2\Delta_z^2\cos^2\theta}}.
\end{multline}

It is instructive to compare the result Eq. (\ref{result}) with corresponding expression from Ref. \onlinecite{Dery1} for
$p=1$, which reads

\begin{equation}
\label{Deryresult}
I^{N\rightarrow F}=\Bigl(\frac{2e}{\hbar}\Bigr)\frac{\Gamma_F\Gamma_N\Delta_z^2\sin^2\theta}
{\big[(2\Gamma_N+\Gamma_F)\Delta_z^2+2\Gamma_F^{2}\Gamma_N\big]-\Gamma_F\Delta_z^2\cos^2\theta}.
\end{equation}
At small $\theta$ we can expand the square root in Eq. (\ref{result}) as
\begin{equation}
\label{expansion}
\sqrt{(\Delta_z^2-\Gamma_F^2)^2+4\Gamma_F^2\Delta_z^2\cos^2\theta}\approx
\Delta_z^2+\Gamma_F^2+\frac{2\Delta_z^2\Gamma_F^2\theta^2}{\Delta_z^2+\Gamma_F^2}.
\end{equation}
It follows from Eq. (\ref{expansion})
that the two results coincide at small $\theta$.  Otherwise, they are different, see Fig. \ref{6}. The difference is most pronounced for
 $\Gamma_F \ll \Gamma_N$, when the tunneling into $F$ dominates the current.
For example, for particular values $\Delta_z=\Gamma_F$ and $\theta=\pi/2$, the current Eq. (\ref{result}) is two times bigger
than $I^{N\rightarrow F}$ given by Eq. (\ref{Deryresult}).
The origin of the discrepancy is the form of the Hamiltonian, adopted in Ref. \onlinecite{Dery1}, where strong Coulomb repulsion is ascribed to electrons in the states $\uparrow$ and $\downarrow$. This is permissible only for $\theta=0$.
For nonzero $\theta$, the repulsion takes place between the electrons occupying the eigenstates $\chi_+$ and $\chi_-$, see
Eq. (\ref{spinors}). Comparison of Eqs. (\ref{result}) and (\ref{Deryresult}) is presented in Fig. \ref{6}.

At large $\Delta_z$ the current Eq. (\ref{result}) saturates at the value
\begin{equation}
\label{infiniteu}
I^{N\rightarrow F}_\infty=\frac{2e}{\hbar}\frac{\Gamma_F\Gamma_N\sin^2\theta}{\Gamma_F\sin^2\theta+2\Gamma_N}.
\end{equation}
In this limit, the coupling between the Zeeman levels is negligible, so that the value $I^{N\rightarrow F}_\infty$ \hspace{1.5mm}
follows from Eq.~(\ref{TwoCurrents}), with $\tau^{N\rightarrow i}=1/2\Gamma_N$ and $\tau^{i\rightarrow F}=1/\Gamma_F\sin^2
\theta$. Naturally, the large-$\Delta_z$ limit of Eq. (\ref{Deryresult}), in which the coupling of the Zeeman levels is completely neglected, coincides with Eq. (\ref{infiniteu}).

In closing of this Section we present the expression for the current  which generalizes Eq. (\ref{result}) to the case of
a finite polarization of the ferromagnetic electrode

\begin{widetext}
\begin{equation}
\label{FinitePolarization}
I^{N\rightarrow F}= \Bigl(\frac{2e}{\hbar}\Bigr) \frac{\Gamma_N\big(\Delta_z^2+(2-p^2)\Gamma_F^2-\sqrt{(\Delta_z^2-p^2\Gamma_F^2)^2+4p^2\Gamma_F^2\Delta_z^2\cos^2\theta}~\big)}
{4\Gamma_N\Gamma_F+\Delta_z^2+(2-p^2)\Gamma_F^2-\sqrt{(\Delta_z^2-p^2\Gamma_F^2)^2+4p^2\Gamma_F^2\Delta_z^2\cos^2\theta}}.
\end{equation}
\end{widetext}

\begin{figure}
\includegraphics[width=80mm]{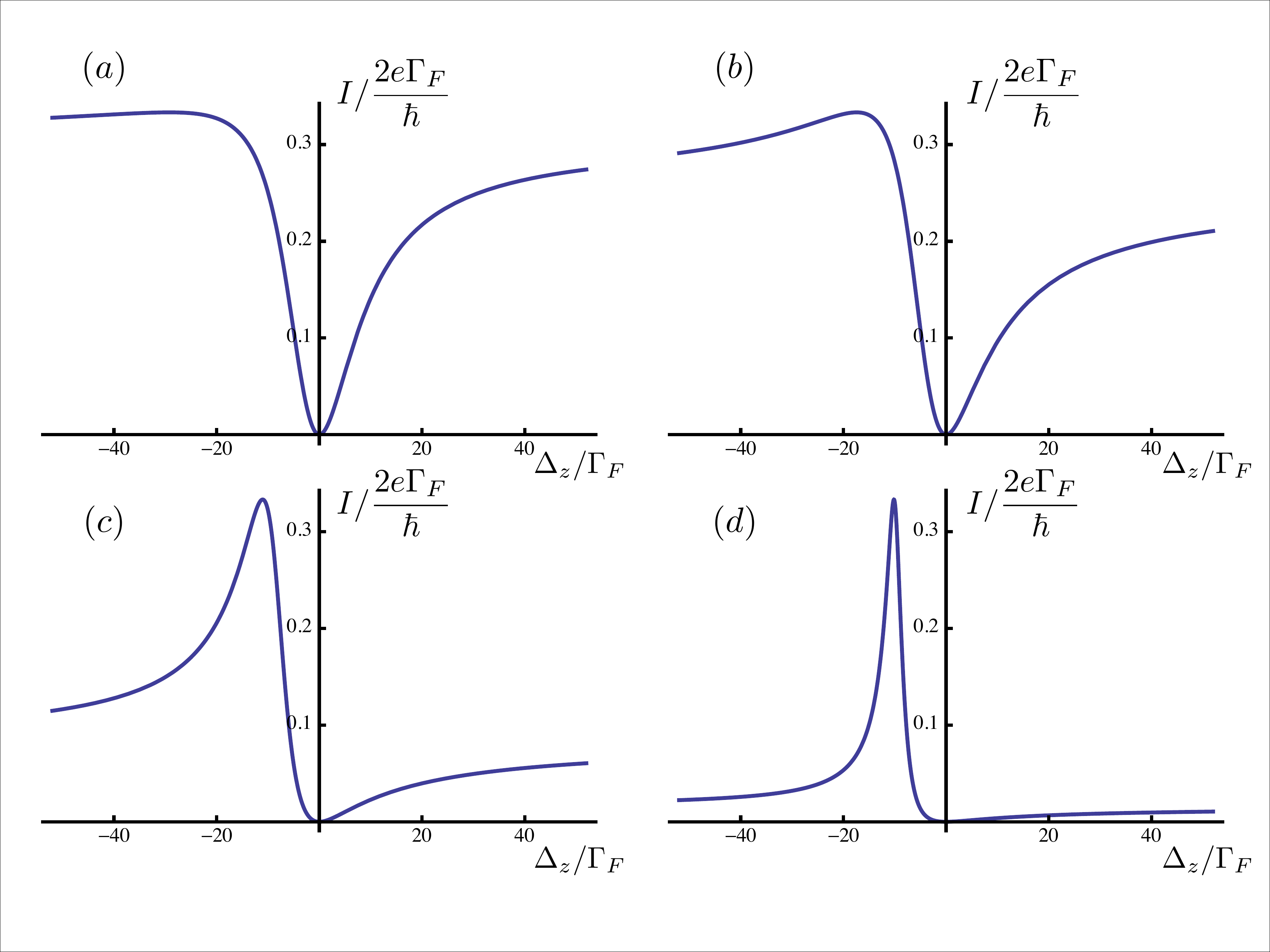}
\caption{The effect of pseudomagnetic field on magnetotunneling. The current (in the units $2e\Gamma_F/\hbar$) in the correlated regime is plotted versus dimensionless
magnetic field, $\Delta_z/\Gamma_F$, for orientations $\theta =70^\circ$ (a), $\theta =55^\circ$ (b), $\theta= 25^\circ$ (c), and
$\theta=10^\circ$ (d). The plots correspond to $\Gamma_N=\Gamma_F$ and pseudomagnetic field $\Delta_0=10\Gamma_F$.
 }
\label{7}
\end{figure}

\section{Concluding remarks}

\begin{itemize}

\item

Our main physical message is that in resonant magneto-tunneling between the normal electrode and
the ferromagnet, the effect of coupling of Zeeman levels via a ferromagnetic electrode affects the
current both in correlated and non-correlated regimes. At this point we would like to draw a link to the
earlier studies,  Refs. \onlinecite{SchultzVonOppen}, \onlinecite{Brandes}, where the correlated resonant transport between
the {\em normal} electrodes via a {\em two-level} system, e.g. two quantum dots in parallel\cite{Brandes},  was studied. The authors realized that the current
is strongly affected by the coupling between the levels via continuum of the states in the electrodes, and that
the rate-equations-based description is invalid due to this coupling. They demonstrated that this coupling gives rise
to a strong dependence of current on the energy separation of the levels. In our situation, this separation is simply the Zeeman
splitting, $\Delta_z$.

 In Refs. \onlinecite{SchultzVonOppen}, \onlinecite{Brandes}, the ferromagnet was mimicked by the asymmetry of coupling of the components of the two-level
system to the electrodes. In our situation, the source of asymmetry is the angle, $\theta$, between the magnetic field and the
magnetization.
The effect analogous to ``magnetoresistance" was captured in Refs. \onlinecite{SchultzVonOppen}, \onlinecite{Brandes} by numerically solving the master equations. Our situation, when only one electrode is ferromagnetic, is simpler,
which allowed us to get the analytical result Eq. (\ref{result}) for the correlated current.

\item

In the correlated regime, the magnetoresistance is present only for one current direction $N\rightarrow F$.
Our result Eq. (\ref{current}) suggests that outside the blockaded regime $V>U$,
when the current is the same for both voltage polarities,
the magnetoresistance is still finite and strong. Probably, this prediction, equal mangetoresistances for both bias polarities,
 for high enough bias can be tested in 3T spin-transport experiments.

\item

Except for the papers Refs. \onlinecite{Martinek-1}, \onlinecite{Braig}, the bulk of theoretical studies\cite{Martinek-2,Martinek-1,Braun,Martinek0,Martinek2,Martinek3,Martinek4,Tserkovnyak}
of resonant transport between two ferromagnetic electrodes was focused on the low-temperature Kondo regime.
As it was pointed in Ref. \onlinecite{Martinek-1}, outside the Kondo regime, in addition to blocking,  there is
another  prominent many-body effect, which results from the polarization of the electrodes and might affect the
transport. Namely,  the on-site repulsion gives rise to a pseudomagnetic field
\begin{equation}
\label{pseudomagnetic}
\Delta_0=\frac{\Gamma_F}{\pi}\int_{-V/2}^{V/2}d\varepsilon \Bigl(\frac{1}{\varepsilon-U}-\frac{1}{\varepsilon}\Bigr)
\end{equation}
directed along the magnetization. The structure of Eq. (\ref{pseudomagnetic}) suggests that the underlying mechanism is
similar to cotunneling. Incorporating of this field into Eq. (\ref{result}) is performed by replacing $\Delta_z\cos\theta$
with $\Delta_z\cos\theta +\Delta_0$ and $\Delta_z^2$ with $\Delta_z^2+\Delta_0^2+2\Delta_z\Delta_0\cos\theta$. The effect
of pseudomagnetic field on the shape of magnetoresistance curves is illustrated in Fig. \ref{7}.
We see that for large enough $\Delta_0 \sim 10\Gamma_F$ the shapes can undergo a dramatic transformation
becoming asymmetric and even non-monotonic. Still these shapes do not explain experimental observation that the
current grows with $\Delta_z$  at $\theta=\pi/2$ and drops with $\Delta_z$  at $\theta =0$. To account for this observation
it was assumed in Ref. \onlinecite{Dery1} that, in addition to external field, a strong in-plane hyperfine field is present.

\item

Throughout the paper we assumed that the impurity level position, ${\cal E}_0$, is zero. In fact, we required that ${\cal E}_0$ lies within the interval $(-\frac{V}{2}, \frac{V}{2})$, see Fig.~\ref{1}.
For  ${\cal E}_0$ lower than $-\frac{V}{2}$ the resonant tunneling is forbidden. It will be allowed again\cite{GlazmanMatveev2} when ${\cal E}_0$ falls into
the interval $(-\frac{V}{2}-U, \frac{V}{2}-U)$ (impurity of the ``type B" in the language of Ref. \onlinecite{Dery1}). Then the intermediate state for the passage of current will be doubly occupied, and magnetoresistance will be present\cite{Dery1} for  $I^{F\rightarrow N}$, but absent for $I^{N\rightarrow F}$.
If ${\cal E}_0$ is lower than $-\frac{V}{2}$ but above $\frac{V}{2}-U$,  the mechanism of passage of current is cotunneling,
 i.e. an elastic two-electron process in course of which one electron leaves the impurity to $N$ and another arrives from $F$. The cotunneling  rate, $\tau_c^{-1}$, is given by $\tau_c^{-1}\sim \Gamma_F\Gamma_N/{\cal E}_0$, so that the magnitude of current is $I_c=e/\tau_c$. There is a
question whether or not the cotunneling current,  $I^{F\rightarrow N}$,   exhibits the magnetic field dependence. In our opinion it does. Indeed, without the magnetic field and  for fully polarized $F$ electrode, the state of the impurity after a single cotunneling act is $\uparrow$. This forbids the next cotunneling act, so that $I_c=0$. Finite magnetic field lifts this blockade in the same way as it does for a direct resonant  current. We thus expect the mangetoresistance of the form $I^{F\rightarrow N}(\Delta_z)=e\Bigl[\Delta_z^2\tau_c\sin^2\theta/\left(1+\Delta_z^2\tau_c^2\right)\Bigr]$.

\end{itemize}

\section{Acknowledgements}
We are grateful to M. C. Prestgard and A. Tiwari for piquing our interest in 3T spin transport.
 This work was supported by NSF through MRSEC DMR-1121252.

\end{document}